\documentclass[final,times,5p,xtwocolumn]{elsarticle}

\usepackage{amssymb}
\usepackage{amsmath}
\usepackage{latexsym,epsfig}
\usepackage{graphicx}
\usepackage{hyperref}

\usepackage{amsfonts}
\usepackage{epsfig}
\usepackage{dcolumn}
\usepackage{bm}

\newcommand{\ba}{\begin{array}}
\newcommand{\ea}{\end{array}}
\def\br{\begin{eqnarray}}
\def\er{\end{eqnarray}}
\def\be{\begin{equation}}
\def\ee{\end{equation}}

\def\({\left(}
\def\){\right)}

\journal{Physics Letters B}

\begin{document}

\begin{frontmatter}



\title{Non-perturbative fixed points and renormalization group improved effective potential}


\author[a1]{A. G. Dias}
\author[a1]{J. D. Gomez}
\author[a1,a2]{A. A. Natale}
\author[a1]{A. G. Quinto}
\author[a1]{A. F. Ferrari}

\address[a1]{Universidade Federal do ABC, Centro de Ci\^encias Naturais e Humanas, Rua Santa Ad\'elia, 166, 09210-170, Santo Andr\'e, SP, Brasil}
\address[a2]{Instituto de F\'{\i}sica Te\'orica, UNESP, Rua Dr. Bento T. Ferraz, 271, Bloco II, 01140-070, S\~ao Paulo, SP, Brazil}


\begin{abstract}
The stability conditions of a renormalization group improved effective potential have been discussed
in the case of scalar QED and QCD with a colorless scalar. We calculate the same potential in these models assuming the existence of
non-perturbative fixed points associated with a conformal phase. In the case of scalar QED the barrier of instability found previously is
barely displaced as we approach the fixed point, and in the case of QCD with a colorless scalar not only the barrier is changed but the local
minimum of the potential is also changed.
\end{abstract}


\end{frontmatter}


\section{Introduction}
\label{intro}

The discovery of a Higgs-like particle at the CERN-LHC, and the fact that this particle is ``lighter" than what could be expected for the Higgs boson in several extensions of the Standard Model (SM) is leading to a deeper investigation of the mass generation mechanism as it is known in the SM. Many recent papers are discussing the Higgs mechanism under new points of view, such as the naturality of the model \cite{n1,n2,n3}, its stability \cite{s1,s2,s3}, and studying possible alternatives or extensions of the model.

Several years ago the possibility that a conformal classical symmetry could be important in the mass generation mechanism was discussed by Meissner and Nicolai \cite{mn}. At that time they proposed an extension of the SM where the radiative symmetry breaking calculated with the help of the effective potential, as first suggested by Coleman and Weinberg (CW) \cite{cw}, was compatible with the experimental data then available. The example of \cite{mn} was already giving an answer to the questions of naturality and stability of the Higgs mechanism, and the possibility that the SM symmetry breaking could be implemented through radiative corrections is still in discussion \cite{hill}.

How the CW effective potential calculation is applicable and reliable in a realistic theory is a motive of debate. Meissner and Nicolai discussed the applicability of a renormalization group (RG) improved version of the one-loop CW effective potential in simple models \cite{mn2}, where the behavior of the coupling constants could be easily calculated. The inclusion of the coupling constants evolution extends the validity range of the effective potential. One of the criteria for applicability of the RG improved CW effective potential proposed in \cite{mn2} was that the running coupling constants, expressed as functions of the classical field, should stay small. Of course, away from the origin the RG $\beta$ functions will depend on the renormalization scheme
and how the coupling is defined. However, we should expect the stability of the effective potential certainly to depend on the coupling constants RG behavior in a much larger range of values \cite{lattice}.

In the examples of classically conformal theories of \cite{mn2} the effective potential stability is connected to
ultraviolet (UV) and infrared (IR) barriers caused by the presence of a Landau pole in the QED or QCD couplings. However, the existence of a Landau pole in these couplings has been questioned, and instead of a pole they may present a non-perturbative fixed point. Due to the phenomenon of dynamical symmetry breaking in QED and QCD, the coupling constants may freeze after they reach a certain critical value, whereas in the QCD case, as will be discussed later, such critical value is even not so large. It is the effect of non-perturbative fixed points of these types in the RG improved effective potential calculation, applied to the models of \cite{mn2}, that we want to discuss in this work. Their effect has not been discussed in the context of the CW potential and they may even modify the potential stability conditions.

In QED the  non-perturbative fixed point that we referred to above implies a critical coupling $\alpha_{c}\approx \pi/3$ \cite{mi,ba}, where $\alpha_c$ is the UV critical value of the fine structure constant ($\alpha \equiv e^2/4\pi$). This behavior is a consequence of dynamical chiral symmetry breaking, in a mechanism similar to the fall into the Coulomb center for large charge \cite{mi2}, with a $\beta$ function that is approximated by
\be
\beta_\alpha = - 2 (\alpha - \alpha_c ) \, .
\label{eq0}
\ee
It is not clear whether this fixed point indeed exists, and QED already does not make sense at the physical scale of such critical value, which happens to be above the Planck scale. Nevertheless, the study of such possibility can be instructive.

On the other hand, there are many plausible evidences that QCD develops an infrared non-perturbative fixed point. For instance, the study of dynamical mass generation in QCD indicates that the coupling constant may freeze in the infrared as \cite{cornwall,bjc}
\be
{\bar{g}}^2(k^2)= \frac{1}{\beta_0 \ln[(k^2+4m_g^2)/\Lambda^2]} \, ,
\label{eq1}
\ee
where $\beta_0=(11N-2n_q)/48\pi^2$ with $n_q$ quark flavors, $\Lambda$ is the characteristic QCD scale, and $m_g$ is a dynamically generated
``effective mass" for the gluon, whose preferred value is $m_g \approx 2\Lambda $ \cite{cornwall,natale}. The IR value of Eq. (\ref{eq1}) as
well as the compilation of other IR values for the QCD coupling obtained in different phenomenological applications can be seen in \cite{natale2},
and they do not indicate an abrupt transition to the non-perturbative regime. We can also quote theoretical estimates of $\alpha_s(0)$ through
the functional Schr\"odinger equation, which suggest $\alpha_s(0) \approx 0.5$ \cite{corn1}.
The infrared finite effective charge of QCD, in the context of Schwinger-Dyson equations,
has also been  discussed in \cite{agui} and is associated with an infrared finite gluon propagator.
Actually, finite IR gluon propagators have been confirmed in lattice simulations \cite{cucchieri,bogo}, and they do lead to a non-perturbative IR fixed point \cite{natale3}. The effect of such non-perturbative coupling constant has not been explored in the case of a CW potential calculation
involving QCD.

The organization of this work is the following: In Sec. II we discuss the RG improved potential of scalar QED in the presence of a non-perturbative fixed point. This section is a simple example of what we will calculate in the more elaborated case of the CW potential for QCD with a colorless scalar, which is going to be presented in Sec. III. In Sec. IV we draw our conclusions.

\section{CW potential in scalar QED with a fixed point}

The RG improved effective potential of \cite{mn2} for an ordinary scalar field $\varphi$ theory with quartic self-interaction and no classical mass term is given by
\be
W_{eff} (\varphi,g,v)={\hat{g}}_1(L)\varphi^4 {\rm exp}\left[2\int_{0}^{L} {\bar{\gamma}}({\hat{g}}(t)) dt\right] \, ,
\label{eq2}
\ee
where ${\hat{g}}$ may indicate a set of coupling constants, $v$ is some renormalization mass scale,
\be
L\equiv \ln\frac{\varphi^2}{v^2} \, ,
\label{eq3}
\ee
and ${\bar{\gamma}}({\hat{g}})$ is an anomalous dimension associated with the coupling constants.

We will discuss the effective
potential of Eq. (\ref{eq2}) in the case of massless scalar QED.
The scalar self-coupling and the gauge coupling are respectively given by
\be
y=\frac{g}{4\pi^2},\qquad u=\frac{e^2}{4\pi^2},
\ee
and their RG equations are
\be
2\frac{d\,y}{dL}=a_1 y^2-a_2 yu+a_3u^2, \qquad  2\frac{d\,u}{dL}=2bu^2,
\label{2x}
\ee
where $a_1=5/6$, $a_2=3$, $a_3=9$ and $b=1/12$. The anomalous dimension is
$\gamma(y,u)=cu$, with $c=3/4$. The solutions of Eq. (\ref{2x}) are

\begin{equation}
u(L)=\frac{u_0}{1-bu_0 L},
\label{eq41}
\end{equation}
\br
y(L)&=&\frac{(a_2+2b)}{2a_1}u(L) \nonumber \\
&+&\frac{Au(L)}{2a_1}\tan\Big({\frac{A}{8b}\ln{u(L)+C}}\Big),
\label{eq42}
\er
where
\be
A=\sqrt{4a_1 a_3 - (a_2+2b)^2},
\label{eq5a}
\ee
is a positive quantity and $C$ is a constant chosen to satisfy $y(0)=y_0$.

The RG improved effective potential at one loop is
\begin{equation}
 W_{eff}=\frac{\pi^2 \varphi^4 y(L)}{(1-bu_0 L)^{2c/b}}.
\label{eq5}
\end{equation}
Eq. (\ref{eq5}) is the result obtained in \cite{mn2}. This result has one striking difference
in relation to the unimproved potential, which is the presence of two barriers, one related to the
UV Landau pole and an IR one where the potential becomes unbounded from below.

\begin{figure}
\setlength{\epsfxsize}{0.9\hsize} \centerline{\epsfbox{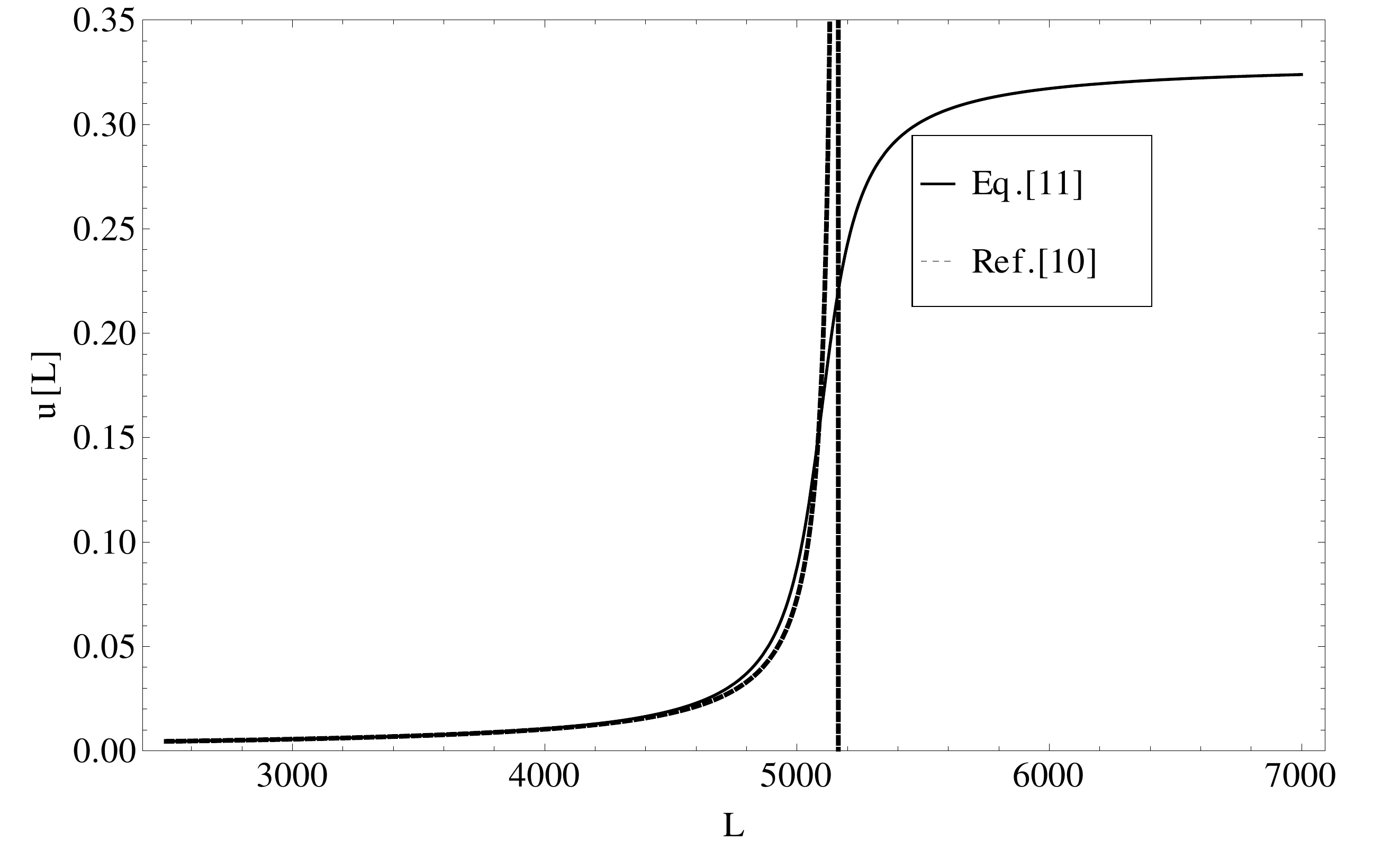}}
\caption[dummy0]{Evolution of the QED gauge coupling giving by Eq. (\ref{eq41}) (dashed curve)  and the fit of Eq. (\ref{eq6})
(continuous curve), which assumes the existence of a non-perturbative fixed point. For small $L$ values the fit agrees with the perturbative result.} \label{fig:QED}
\end{figure}

Let us now suppose that the theory has a fixed point at
$\alpha_{c} (L) \approx 1$. Note that this would be a possibility for QED with fermions as discussed in Refs.\cite{mi,ba}, but here this is just an
{\it{ad hoc}} supposition to exemplify what may happen with the effective potential in the case of a possible freezing of
the coupling constant. We see in Fig. (\ref{fig:QED}) that the critical point of the perturbative coupling will occur for $L>5000$, while for much smaller $L$ values the coupling follows
Eq. (\ref{eq41}). As a consequence, for small values of $L$ we have $u=e^2/4\pi^2 $, and at large $L$ the coupling freezes at $u\approx 1/\pi$.

The main difference between this work and previous calculations of the RG improved effective potential is the introduction of an interpolating coupling constant
joining the perturbative to the non-perturbative regime. The best interpolation formula
between these values is given by a ${\rm tan}^{-1}$ function. We make a fit for the gauge coupling assuming the RG standard solution for $L<4500$, and interpolate it in the region
$4500<L<5300$ with a ${\rm tan}^{-1}$ formula, such that it will be joined to the frozen value of the coupling for $L>5300$. A fit that is reasonable
from $L=0$ up to $L<4000$ at $0.1 \%$ level is given by
\br
 u_{Fit}(L)&=& 0.105 \tan^{-1}\Big(\frac{L-5100}{110}\Big) \nonumber \\
&+& 0.165.
\label{eq6}
\er
In Fig. (\ref{fig:QED}) we show the gauge coupling as a function of $L$, where the dashed curve is the ordinary perturbative behavior with the pole and the continuous curve is
the one showing a freezing coupling constant above the critical QED coupling.

\begin{figure}
\setlength{\epsfxsize}{0.9\hsize} \centerline{\epsfbox{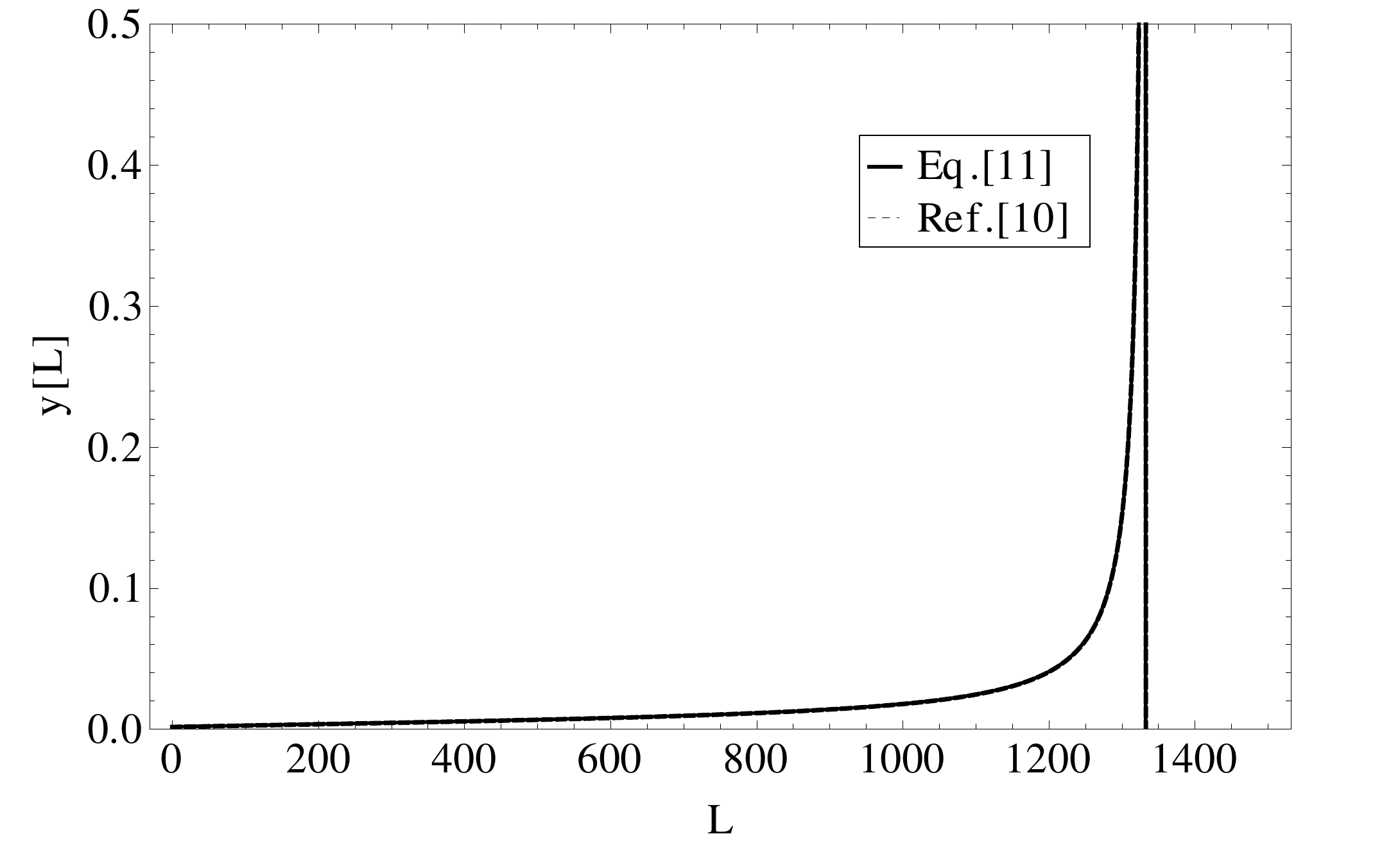}}
\caption[dummy0]{Evolution of the scalar self-coupling. The continuous curve is the result that could be obtained following \cite{mn2}, and
the dashed curve is the result obtained with the existence of the UV frozen QED gauge coupling constant.} \label{fig:QEDcoupling:y}
\end{figure}

There is no pole associated with the gauge coupling, and with the help of Eq. (\ref{eq6}) we can numerically compute the RG solution for the scalar self-coupling shown in Eq. (\ref{eq42}), which still has an IR pole. The behavior
of this coupling is shown in Fig. (\ref{fig:QEDcoupling:y}). In this figure it is possible to see that the barrier present in the purely perturbative calculation is barely changed, and this is the point where the potential will become unbounded from below \cite{cw}.

\begin{figure}
\setlength{\epsfxsize}{0.9\hsize} \centerline{\epsfbox{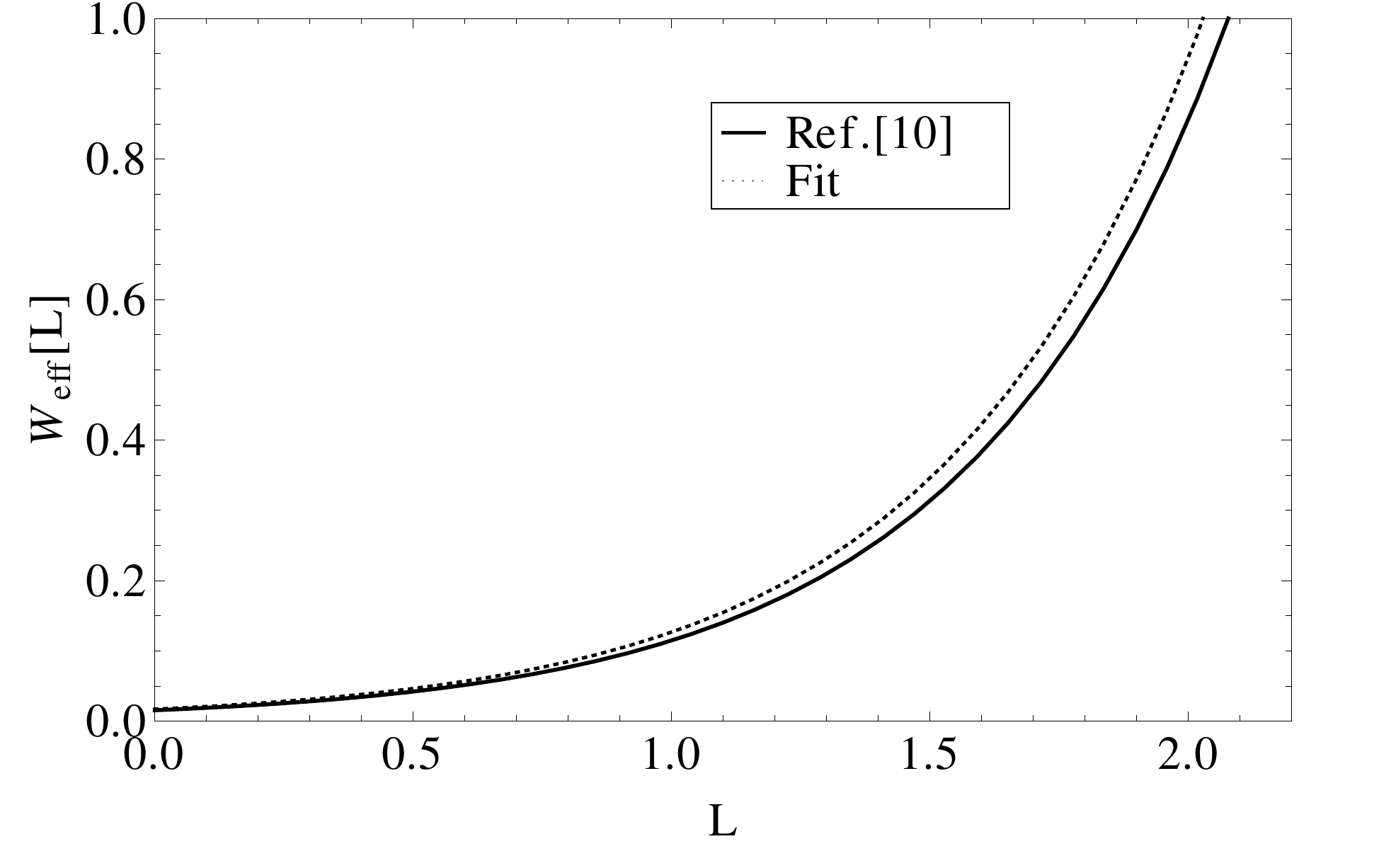}}
\caption[dummy0]{RG group improved effective potential for scalar QED. The continuous curve is the result according to \cite{mn2} and the
dashed curve is the result obtained with the gauge coupling given by Eq. (\ref{eq6}).} \label{fig:WQED}
\end{figure}

The behavior of the effective potential is slightly changed for small $L$ as may be seen in Fig. (\ref{fig:WQED}). Note that in the QED case the minimum
of the effective potential is located in a region of small values of the gauge coupling constant, therefore it is not affected by the non-perturbative UV
fixed point. A more interesting case, where the effective potential may change considerably, will be shown in the sequence.

\section{QCD with a colorless scalar}

Let us now consider the following Lagrangian:
\br
{\cal{L}}&=& -\frac{1}{4}F_{\mu\nu}F^{\mu\nu}+{\bar{q}}\gamma^\mu D_\mu q-\frac{1}{2}\partial_\mu \phi\partial^\mu \phi \nonumber \\
&+&g_Y\phi{\bar{q}}q-\frac{\lambda}{4}\phi^4 \, ,
\label{eq7}
\er
where $q$ are quark fields, $\phi$ is a colorless scalar and $F^{\mu\nu}$ is the usual QCD field tensor. In this case we have
three different coupling constants: a) The gauge coupling $g_s$ that appears in $F^{\mu\nu}$ and in the covariant derivative $D_\mu$, which,
for convenience, will be redefined in the following as
\be
z\equiv \frac{g_s^2}{4\pi^2} \equiv \frac{\alpha_s}{\pi},
\ee
b) the Yukawa coupling $g_Y$, redefined as
\be
x\equiv \frac{g^2_Y}{4\pi^2} ,
\ee
and c) a scalar self-coupling $\lambda$, which will be referred as
\be
y\equiv \frac{\lambda}{4\pi^2}.
\ee
The RG equations obeyed by these couplings at one-loop are
\begin{gather}
  2\frac{d\,y}{dL}=a_1 y^2+a_2 yx-a_3x^2, \label{eq:ydif}\\
  2\frac{d\,x}{dL}=b_1x^2-b_2xz,  \label{eq:xdif}\\
  2\frac{d\,z}{dL}=-cz^2 \label{eq:zdif},
\end{gather}
with the scalar anomalous dimension given by $\gamma(x,y,z)=-hx$, where the values of the different parameters appearing in these equations are
$a_1=6$, $a_2=3$, $a_3=3/2$, $b_1=9/4$, $b_2=4$, $c=7/2$ and $h=3/4$ \cite{sher}.

One can find particular solutions for $z(L)$ and $x(L)$ as follows
\begin{equation}
 \begin{gathered}
    z(L)=\frac{z_0}{1+cz_0 L/2},\\
    x(L)=\frac{b_2-c}{b_1-Kz(L)^{1-b_2/c}}z(L),
    \label{eq:solutionQCD}
 \end{gathered}
\end{equation}
where $K$ is chosen to satisfy $x(0)=x_0$. The parameter $z_0$ is found in order to obtain a coupling constant
compatible with physical values of the QCD coupling. For instance, typical values in \cite{mn2} were
assumed to be $x_0=0.120$ and $z_0=0.249$.
The $x(L)$ can be also written as
$$x(L)=\frac{b_2-c}{b_1}z(L)\Big(1-\frac{K}{b_1}z(L)^{1-b_2/c}\Big)^{-1}=\\$$
$$=\frac{b_2-c}{b_1}z(L)\Big(1+\frac{K}{b_1}z(L)^{1-b_2/c}+\cdots\Big),$$
and since $b_2>c$ the power of $z(L)$ in the denominator will be negative and at the weak
coupling limit $x(L)$ has the following behavior
\be
x(L)\rightarrow\beta z(L),
\ee
where
\be
\beta=\frac{b_2-c}{b_1}.
\ee
With the $x(L)$ solution and the differential equation (\ref{eq:ydif}), we obtain $y(L)$ at the weak coupling limit,
\be
y(L)=\rho z(L),
\label{eqx}
\ee
where $\rho$ is a constant
and the effective potential is
\begin{equation}
 \label{eq:WQCD}
 W_{eff}(L)=\sqrt{2}\pi^2 y(L)\varphi^4\Bigg(\frac{b_1 z(L)^{b_2/c - 1}-K}{b_1 z_{0}^{b_2/c - 1}-K}\Bigg)^{2h}\rho z(L).
\end{equation}

We now depart from what was done in \cite{mn2} and assume an {\it {ad hoc}}
fixed point like the one shown in Eq. (\ref{eq1}). The QCD IR frozen behavior described in Eq. (\ref{eq1}) was obtained for
pure gauge QCD, and the fixed point of Eq. (\ref{eq1}) is not modified in the presence of a small number of fermions \cite{bjc}. Here we assume that the presence of the colorless scalar boson does not change such behavior. Moreover, since the model is not realistic, the region where the freezing of the coupling constant occurs will be described by an extra parameter, that we will vary in order to see the general behavior of such fixed point.

The QCD coupling constant will be modified in the IR according to
\begin{equation}
 \label{eq:alphaQCD:L}
 z^{np}(L)=\frac{Z^{np}_{0}}{1+A \ln\big({e^{L}+\frac{a^2}{\eta^2}}\big)},
\end{equation}
where $L\equiv \ln (\phi^2 / \eta^2)$.
The parameter $a$ in Eq. (\ref{eq:alphaQCD:L}) plays the same role of the dynamical gluon mass in Eq. (\ref{eq1}),
and $\eta$ is a renormalization scale that can be related to the QCD scale.
Note that the coupling is frozen for $\phi/\eta < a/\eta$, where $a$ is a free parameter in the effective potential
calculation. The change is basically what would be expected in actual QCD with dynamical gluon mass generation.

\begin{figure}
\setlength{\epsfxsize}{1.0\hsize} \centerline{\epsfbox{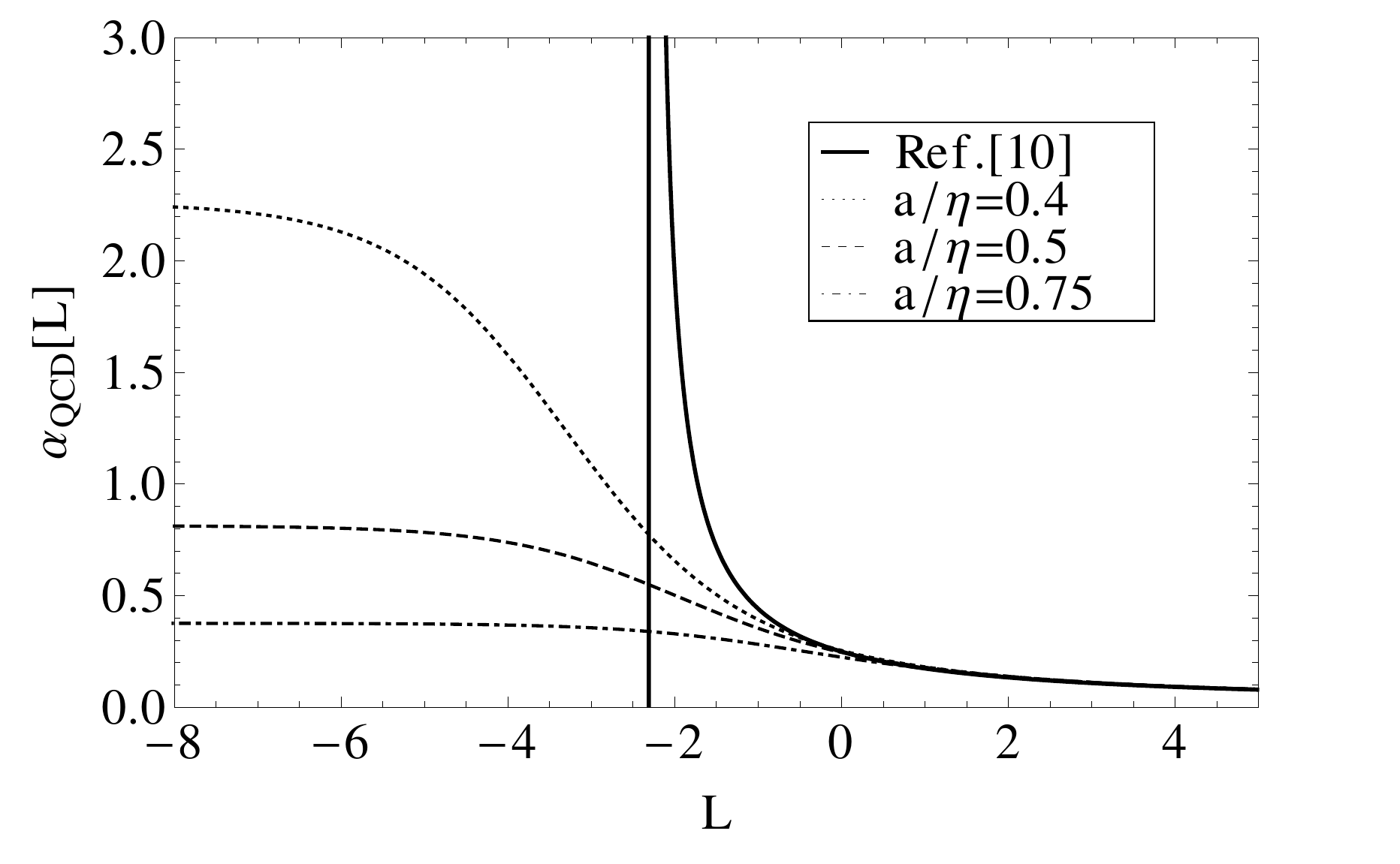}}
\caption[dummy0]{QCD coupling constant for different $\frac{a}{\eta}$ values. The continuous curve shows $z(L)$, i.e. the standard coupling
behavior of \cite{mn2} with the Landau pole at $L\approx -2$, and the discontinuous
         ones show $z^{np}(L)$ as a function of ${a}/{\eta}$.}
 \label{fig:ZnpQCD}
\end{figure}

Consistency between the different couplings implies the same behavior of $z(L)$ and $z^{np}(L)$ for $L\gg 1$ and $\eta\gg 1$
(or ${a^2}/{\eta^2}\ll 1$), i.e. away from the fixed point we must have a perfect match between Eq. (\ref{eq:alphaQCD:L}) and the
perturbative tail of the gauge coupling in Eq. (\ref{eq:solutionQCD}). In order to have this agreement of the couplings in
the perturbative regime we can adjust the $Z^{np}_{0}$ and $A$ values in that region $L>0$.
The values that provide a fit of the $z(L)$ better than $1\%$  are
\be
Z^{np}_{0}=0.272,\qquad A=0.48.
\ee
We can see how Eq. (\ref{eq:alphaQCD:L}) behaves as we change the ratio $\frac{a}{\eta}$
in Fig. (\ref{fig:ZnpQCD}), where it is possible to observe the freezing of the coupling in the infrared region. For comparison,
the result of \cite{mn2} is also shown, appearing in the region where $\frac{a^2}{\eta^2} \ll 1$.

\begin{figure}
\setlength{\epsfxsize}{0.9\hsize} \centerline{\epsfbox{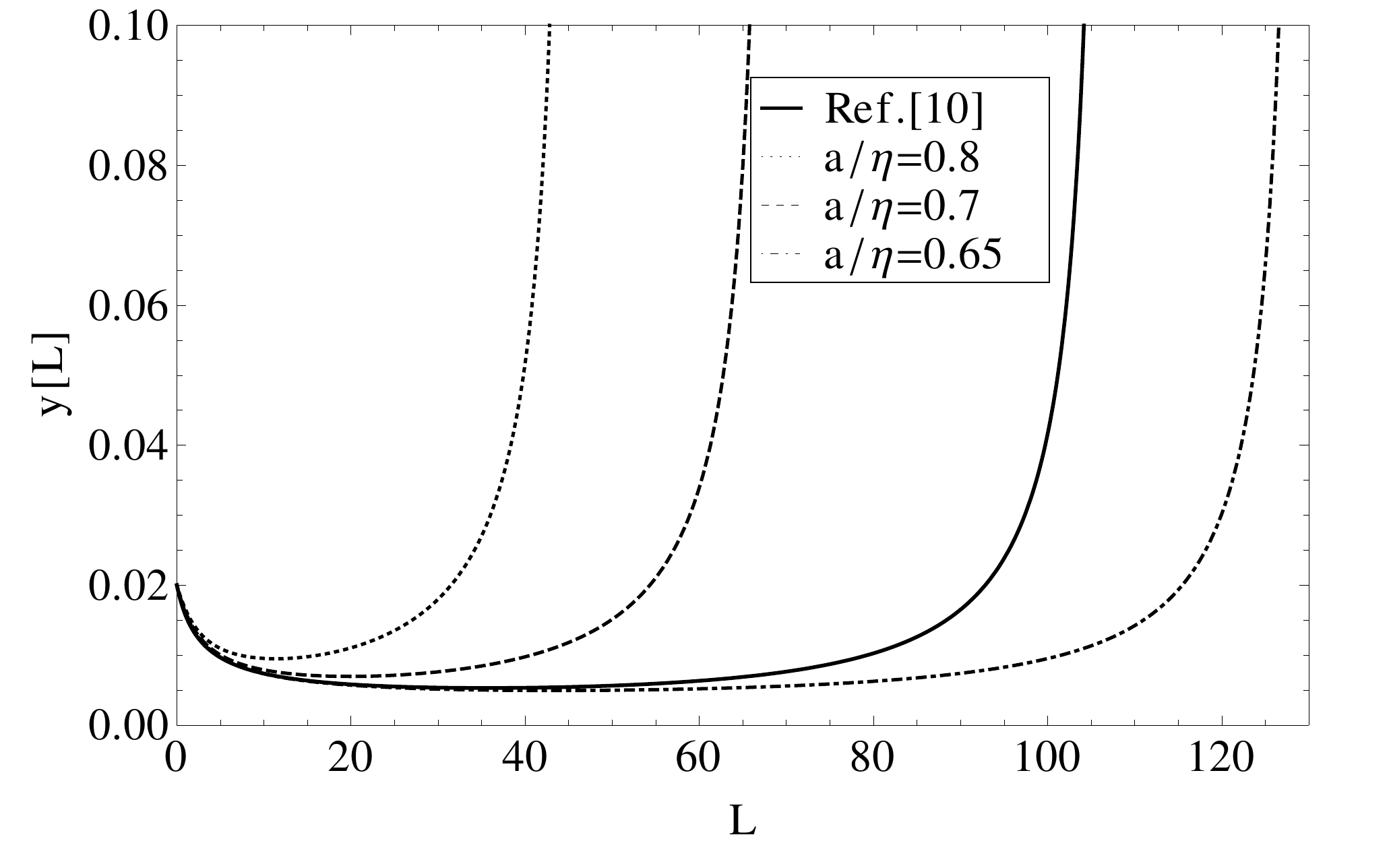}}
\caption[dummy0]{Scalar coupling constant $y(L)$. The continuous bold curve shows $y(L)$ with $a/\eta \cong 0.67$, whereas the other dashed curve
   give $y^{np}(L)$ for different $\frac{a}{\eta}$ values.}
    \label{fig:YnpQCDexata}
  \end{figure}

The scalar coupling $y(L)$ can be obtained in two ways. In the first one we can approximate $x(L)$ in Eq. (\ref{eq:solutionQCD})
as
\be
x(L)\rightarrow\beta z^{np}(L),\qquad \beta=\frac{b_2-c}{b_1},
\ee
then we find
\be
y^{np}(L)\approx C z^{np}(L),
\label{eqy}
\ee
where $C>0$ is a constant.
This result is the equivalent of Eq. (\ref{eqx}), i.e. an approximation for the scalar coupling in the very weak coupling regime.
Therefore the scalar coupling behavior will be similar to the QCD gauge coupling, since the only difference is the
constant $C$. In the second case we consider $x(L)$ given by Eq. (\ref{eq:solutionQCD}) and solve numerically Eq. (\ref{eq:ydif})
obtaining the full non-perturbative behavior of the coupling $y^{np}(L)$. The numerical
results using the same initial conditions of \cite{mn2} ($x_0=0.129$, $y_0=0.020$, $z_0=0.249$) are shown in Fig. (\ref{fig:YnpQCDexata}). We noticed that the scalar coupling is quite dependent on the ratio
$\frac{a^2}{\eta^2}$. The result of \cite{mn2} at large $L$ is obtained with $\frac{a}{\eta}\approx 0.67$ and changes as we move away from
this value.

\begin{figure}
\setlength{\epsfxsize}{1.0\hsize} \centerline{\epsfbox{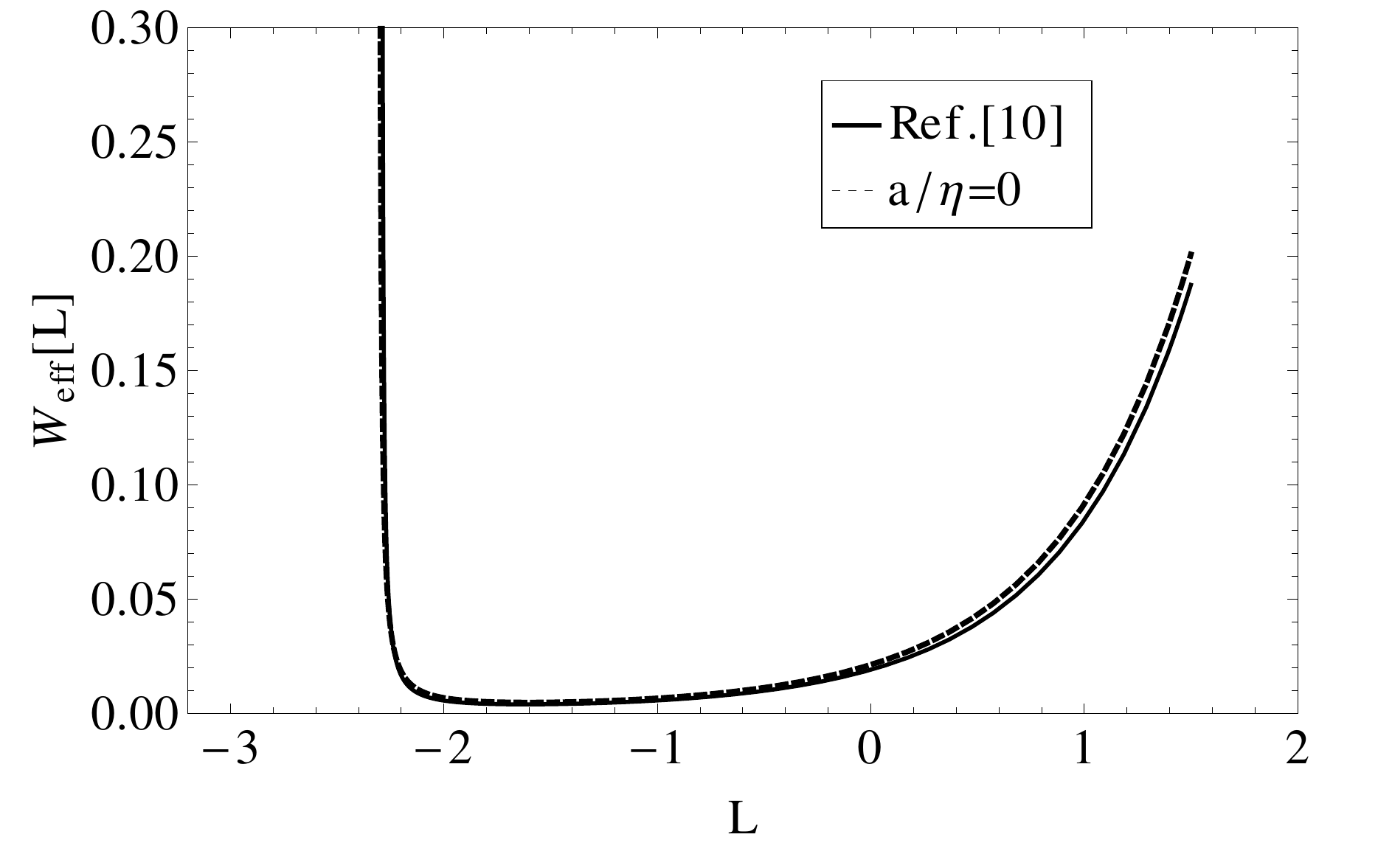}}
\caption[dummy0]{Effective potential of QCD with a colorless scalar.  The bold continuous curve is the result of \cite{mn2},
and the dashed curve is the effective potential obtained with the non-perturbative coupling of Eq. (\ref{eq:alphaQCD:L})
calculated in the limit $\frac{a}{\eta}\approx 0$, i.e. with the analytic result of Eq. (\ref{eqy}) in the small $L$ limit. }
  \label{fig:WQCD}
\end{figure}

\begin{figure}
\setlength{\epsfxsize}{0.9\hsize} \centerline{\epsfbox{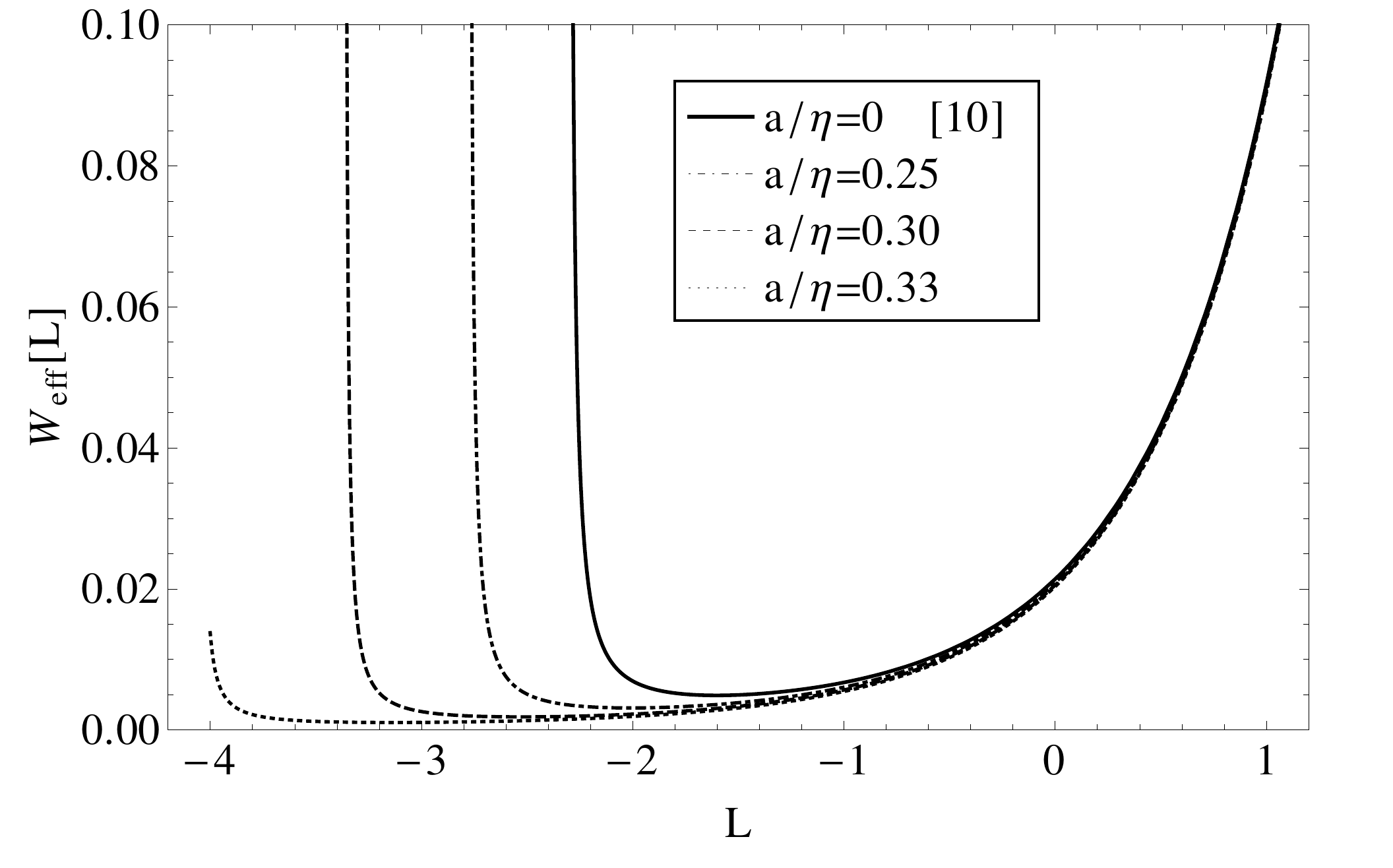}}
\caption[dummy0]{RG improved effective potential of QCD with a colorless scalar. The different curves were obtained considering the non-perturbative coupling of Eq(\ref{eq:alphaQCD:L}) for different values of $\frac{a}{\eta}$, compared to the result of \cite{mn2}.}
  \label{fig:WQCD1}
\end{figure}

Finally we can consider the effective potential given by Eq. (\ref{eq:WQCD}). This one is shown in Fig. (\ref{fig:WQCD})
where it is possible to see that the effective potential of \cite{mn2} at small $L$ is recovered for $\frac{a}{\eta}\approx 0$. However,
this potential has a different behavior as we slowly increase the ratio $\frac{a}{\eta}$ what can be observed in
Fig. (\ref{fig:WQCD1}).
The striking fact is that the presence of the fixed point changes the position of the minimum of the
potential when compared to the one shown in Fig. (\ref{fig:WQCD}).

\begin{figure}
\setlength{\epsfxsize}{1.0\hsize} \centerline{\epsfbox{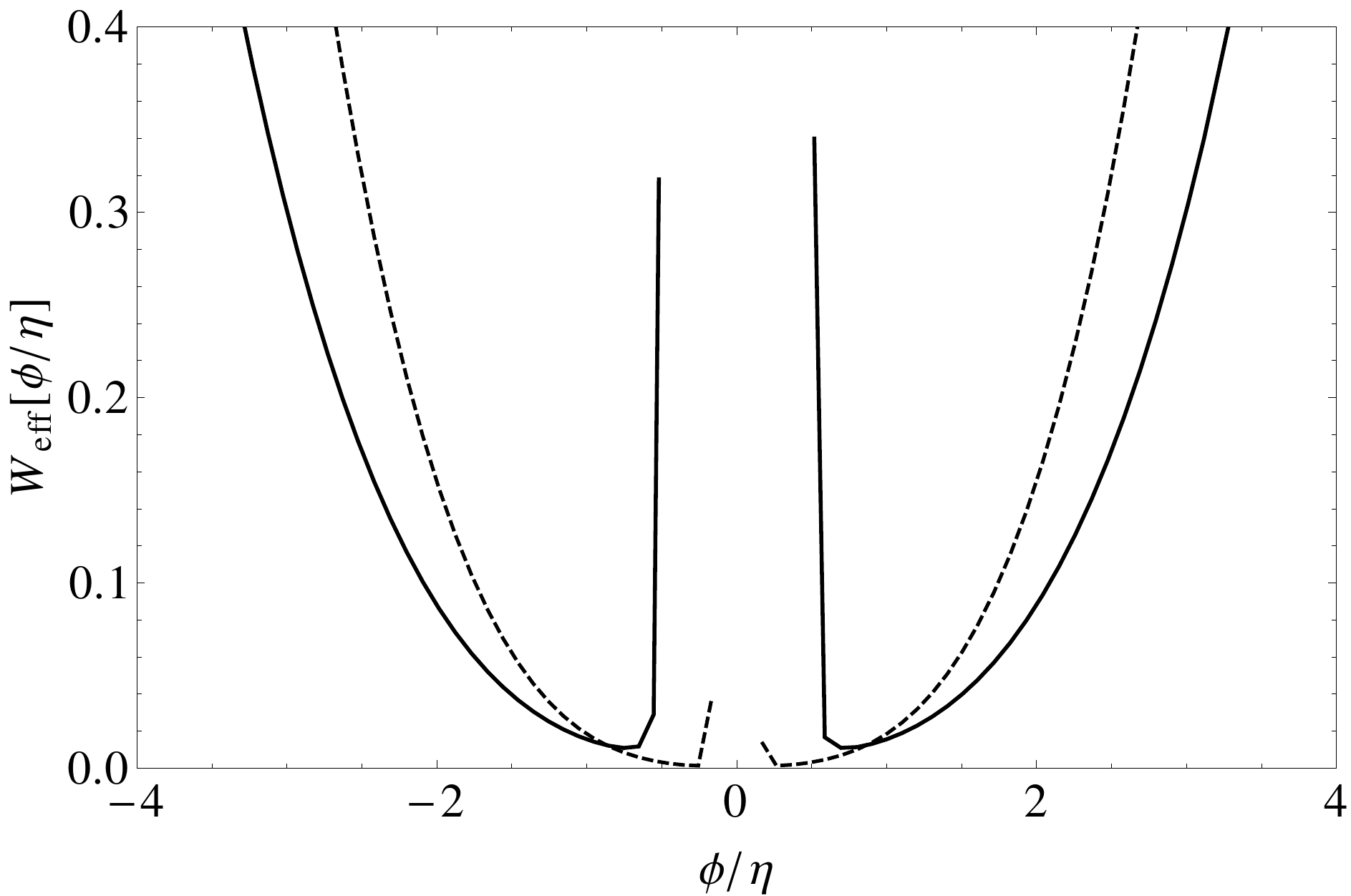}}
\caption[dummy0]{RG improved effective potential of QCD with a colorless scalar as a function of the $\phi$ field. The continuous curve
is the result of \cite{mn2} giving by Eq. (\ref{eq:WQCD})
and the dashed curve is the potential obtained with the non-perturbative couplings ($z^{np}$ and $y^{np}$), where we assumed the value ${a}/{\eta}=0.30$}
  \label{figult}
\end{figure}

\section{Conclusions}

The stability and reliability of the CW potential can be better studied in the RG improved effective potential approach.
However, it is clear that a full analysis of the stability condition of the SM can imply in new physics \cite{mn} as well as on the non-perturbative
analysis of the Higgs potential \cite{lattice}. Here we consider the simple examples of scalar QED and QCD with a colorless scalar in order to verify the effect of possible
non-perturbative fixed points in the RG effective potential calculation. In \cite{mn2} the existence of UV and IR barriers describe the validity
of the effective potential in these models and we verify that the existence of the fixed point we referred to above changes the barriers location.

We consider scalar QED and assume the existence of a conformal behavior at some critical value of the coupling constant. This is a totally \textit{ad hoc}
assumption, however this possibility has been extensively considered in the case of QED with fermions and with possible non-trivial four-fermion interactions \cite{mi,ba,mi2}.
This case is just one example to show the procedure that is going to be applied in the case of QCD with a colorless scalar. The main result is that the barriers
found in \cite{mn2} basically do not change, and the minimum of the potential is not modified, since it happens at small values of the gauge coupling constant,
away from the critical coupling characterizing the non-perturbative fixed point.


Our second example is QCD with a colorless scalar. Here we assume that the QCD gauge coupling has a non-trivial IR fixed point
similar to the one that is expected when QCD exhibits the phenomenon of dynamical gauge boson mass generation \cite{cornwall,bjc,natale3}. In this case
the gauge coupling and scalar coupling barriers found in \cite{mn2} are modified, due to the non-existence of a Landau pole in the IR behavior of the
QCD gauge coupling. However, the most interesting fact is that the minimum value of the potential is changed, as can be observed in Fig. (\ref{figult}), because
in this case the minimum is located in one region of the coupling constant that is governed by the non-perturbative fixed point. Of course, these are
non-realistic examples but they may indicate that such effect may also affect the calculation of the RG improved effective potential in realistic models.
We believe this is a possibility that deserves further study.

\vskip0.5cm
\textbf{Acknowledgments}

This research was  partially supported by the grants 303094/\-2013-3 (A.G.D), 301755/2010-8 (A.A.N), and 482874/2013-9 (A.F.F) of Conselho Nacional de Desenvolvimento Cient\'{\i}fico e Tecnol\'ogico (CNPq); by the grants 2013/22079-8 (A.G.D, A.A.N, and A.F.F) and 2013/24065-4 (J.D.G) of Funda\c{c}\~{a}o de Amparo \`{a} Pesquisa do Estado de S\~ao Paulo (FAPESP); by the grant 076893/2014  (A.A.N) and p.h.d scholarship (A.G.Q) of Coor\-de\-na\-\c c\~ao de Aperfei\-\c co\-amento de Pessoal de N\'{\i}vel Superior (CAPES).





\vskip0.5cm
\textbf{References}

\begin{thebibliography}{99}

\bibitem{n1} O. Antipin, M. Mojaza and F. Sannino, arXiv:1310.0957 [hep-ph].

\bibitem{n2} M. Heikinheimo, A. Racioppi, M. Raidal, C. Spethmann and K. Tuominen, arXiv:1304.7006 [hep-ph].

\bibitem{n3} A. Farzininia, H.-J. He and J. Ren, Phys. Lett. B {\bf 727}, (2013), 141.

\bibitem{s1} O. Antipin, M. Gillioz, J. Krog, E. Molgaard and F. Sannino, JHEP {\bf 1308}, (2013), 034.

\bibitem{s2} H. Gies, C. Gneiting and R. Sondenheimer, Phys. Rev. D {\bf 89}, (2014), 045012.

\bibitem{s3} A. Kobakhidze and A. Spencer-Smith, arXiv:1404.4709 [hep-ph].

\bibitem{mn} K.A. Meissner and H. Nicolai, Phys. Lett. B {\bf 648}, (2007), 312; Phys. Lett. B {\bf 660}, (2008), 260.

\bibitem{cw} S. Coleman and E. Weinberg, Phys. Rev. D {\bf 7}, (1973), 1888.

\bibitem{hill} C.T. Hill, arXiv:1401.4185, [hep-ph].

\bibitem{mn2} K.A. Meissner and H. Nicolai, Acta Phys. Polon. B {\bf 40}, (2009), 2737.

\bibitem{lattice} K. Jansen et al., PoS Confinement X, (2012), 276; J. Bulava et al., PoS Lattice 2012, (2012), 054; P.Hedge et al., arXiv:1310.5922 [hep-lat]; arXiv:1310.6260, [hep-lat].

\bibitem{mi} V.A. Miransky, {\it Dynamical symmetry breaking in Quantum Field Theories}, World Scientific (1994).

\bibitem{ba} W.A. Bardeen, C.N. Leung and S.T. Love, Phys. Rev. Lett. {\bf 56}, (1986), 1230; Nucl. Phys. B {\bf 273}, (1986), 649.

\bibitem{mi2} P.I. Fomin and V.A. Miransky, Phys. Lett. B {\bf 64}, (1976), 166.

\bibitem{cornwall} J.M. Cornwall, Phys. Rev. D {\bf 26}, (1982), 1453.
	
\bibitem{bjc} J.M. Cornwall, J. Papavassiliou and D. Binosi, ``The Pinch Technique and its Applications to Non-Abelian Gauge Theories", Cambridge University Press, 2011.

\bibitem{natale} A.A. Natale, PoS QCD-TNT {\bf 09}, (2009), 031;  F. Halzen, G.I. Krein and A.A. Natale, Phys. Rev. D {\bf 47}, (1993), 295.

\bibitem{natale2} A.C. Aguilar, A. Mihara and A.A. Natale, Phys. Rev. D {\bf 65}, (2002), 054011.

\bibitem{corn1} J. M. Cornwall, \textit{Center vortices, the func\-tional Schrodinger equation, and CSB}, Invited talk at the conference ``Approaches to Quantum Chromodynamics", Oberw\"olz, Austria, September 2008, arXiv:0812.0359 [hep-ph].

\bibitem{agui} A. C. Aguilar, D. Binosi and J. Papavassiliou, JHEP {\bf 1007}, (2010), 002; A. C.
Aguilar, PoS QCD-TNT {\bf 09}, (2009), 001; A. C. Aguilar and J. Papavassiliou, Nucl. Phys. Proc.
Suppl. {\bf 199}, (2010), 172; A. C. Aguilar, D. Binosi, J. Papavassiliou and J. Rodrigues-Quintero,
Phys. Rev. D {\bf 80}, (2009), 085018; A. C. Aguilar, D. Binosi and J. Papavassiliou, PoS LC2008,
(2008), 050; A. C. Aguilar and J. Papavassiliou, J. Phys. Conf. Ser. {\bf 110}, (2008), 022040;
A. C. Aguilar and J. Papavassiliou, AIP Conf. Proc. {\bf 964}, (2007), 312.

\bibitem{cucchieri} A. Cucchieri and T. Mendes, PoS QCD-TNT {\bf 09}, (2009), 031; Phys. Rev. Lett. {\bf 100}, (2008), 241601; Phys. Rev. D {\bf 81}, (2010), 016005.

\bibitem{bogo} I. Bogolubsky, E. Ilgenfritz, M. Muller-Preussker and A. Sternbeck, Phys. Lett. B {\bf 676}, (2009), 69.

\bibitem{natale3}  A.C. Aguilar, A.A. Natale and P.S. Rodrigues da Silva, Phys. Rev. Lett. {\bf 90}, (2003), 152001.

\bibitem{sher} M. Sher, Phys. Rep. {\bf 179}, (1989), 273.


\end{thebibliography}


\end{document}